\documentclass[epj]{svjour}

\usepackage[utf8]{inputenc}
\usepackage[T1]{fontenc}
\usepackage{amsmath}
\usepackage{amsfonts}
\usepackage{amssymb}
\usepackage{accents}
\usepackage{mathrsfs}
\usepackage{graphicx}
\usepackage{cite}
\usepackage[colorlinks,hyperindex,plainpages=false]{hyperref}

\DeclareMathOperator{\pr}{pr}
\newcommand{\dd}{\mathrm{d}}
\newcommand{\DD}{\mathrm{D}}
\newcommand{\lc}[1]{\accentset{\circ}{#1}}
\newcommand{\tp}[1]{\accentset{\bullet}{#1}}
\newcommand{\intprod}{\mathrel{\reflectbox{\rotatebox[origin=c]{180}{$\neg$}}}}

\allowdisplaybreaks

\begin{document}
\title{General cosmological perturbations in teleparallel gravity}
\author{Manuel Hohmann}
\institute{Laboratory of Theoretical Physics, Institute of Physics, University of Tartu, W. Ostwaldi 1, 50411 Tartu, Estonia}
\date{Received: date / Revised version: date}
%
\abstract{
We discuss linear perturbations of the most general class of teleparallel spacetimes with cosmological symmetry, and perform a decomposition of these perturbations into irreducible components. We then study their behavior under gauge transformations, i.e., infinitesimal diffeomorphisms of the background spacetime, and derive a full set of gauge-invariant perturbation variables. Comparing our results with perturbations around the diagonal tetrad corresponding to a spatially flat teleparallel background cosmology, we see the appearance of new terms which depend on the particular choice of a cosmologically symmetric background tetrad. Finally, we demonstrate the application of our findings to teleparallel gravity theories. For convenience, results are displayed both in general and conformal coordinates.
} 

\maketitle

\section{Introduction}\label{sec:intro}
Some of the most prominent open questions in gravity theory arise from observations in cosmology, such as the cosmic microwave background radiation~\cite{Aghanim:2019ame,Aghanim:2018eyx,Aghanim:2018oex,Akrami:2019izv,Akrami:2019bkn,Akrami:2018odb}, the large scale structure~\cite{Ahumada:2019vht,Alam:2020sor}, gravitational waves~\cite{Abbott:2016blz,TheLIGOScientific:2017qsa} and supernovae~\cite{Scolnic:2017caz}. While the latter is well modeled assuming an expanding, homogeneous and isotropic universe, the theoretical description of the former crucially relies on the evolution of perturbations of the homogeneous matter distribution. Understanding matter as the source of gravity, and describing gravity via the geometry of spacetime, these perturbations of the matter distribution are complemented by perturbations of the homogeneous and isotropic spacetime geometry. The mathematical description of such cosmological perturbations is therefore an important tool for studying the modern observations in cosmology.

The theoretical treatment of cosmological perturbations of (pseudo-)Riemannian spacetime geometry, which is the mathematical foundation of the standard formulation of general relativity and the most well-known class of its extensions, has a long-standing history~\cite{Lifshitz:1945du,Lifshitz:1963ps,Hawking:1966qi,Harrison:1967zza}. An important simplification of this cosmological perturbation theory has been achieved by studying how perturbations transform under gauge transformations, i.e., infinitesimal coordinate changes which retain the nature of the spacetime geometry being a small perturbation of a homogeneous and isotropic background. This has led to the construction of gauge-invariant perturbation variables, which describe the physical information contained in the metric perturbations, thus separating them from the arbitrary gauge choice, and which have found wide application~\cite{Bardeen:1980kt,Kodama:1985bj,Mukhanov:1990me,Malik:2008im}.

While the standard formulation of general relativity employs the curvature of the Levi-Civita connection of a Riemannian spacetime as the mediator of the gravitational interaction, also other formulations in terms of the torsion or nonmetricity of a flat connection exist~\cite{BeltranJimenez:2019tjy}. While these formulations are equivalent in the sense that they lead to field equations which possess the same solutions for the metric in all three cases, modifications thereof lead to essentially inequivalent classes of theories. Here we focus on the case of a connection which possesses only torsion, but no curvature or nonmetricity, and which is the basis of teleparallel theories of gravity~\cite{Einstein:1928,Moller:1961,Aldrovandi:2013wha}. In the covariant formulation, which we will adopt here, the dynamical fields which describe the spacetime geometry are a tetrad and a flat, metric-compatible spin connection~\cite{Krssak:2015oua,Krssak:2018ywd}.

Numerous studies of homogeneous and isotropic cosmology in modified teleparallel gravity have been performed; see~\cite{Cai:2015emx} for a review. The theory of cosmological perturbations in the framework of teleparallel geometry, however, is less explored. As for the background cosmology, studies are concentrated on a teleparallel background described by a diagonal tetrad, which induces a spatially flat metric, together with a vanishing spin connection. For this background, gauge-invariant perturbations have been obtained and their dynamics in the general class of \(f(T)\) theories has been studied~\cite{Chen:2010va,Izumi:2012qj,Li:2018ixg,Golovnev:2018wbh,Golovnev:2020aon}, as well as its extension to \(f(T,B)\) theories~\cite{Bahamonde:2020lsm}. Other works focus on particular components of the perturbations, such as the scalar perturbations relevant for structure formation~\cite{Wu:2012hs,Nunes:2018xbm,Casalino:2020kdr} or the transverse, traceless tensor perturbations, which model the propagation of gravitational waves~\cite{Bahamonde:2019ipm}.

While a spatially flat, homogeneous and isotropic teleparallel background geometry can be modeled by a diagonal tetrad and vanishing spin connection, this is not the case for a spatially curved background~\cite{Ferraro:2011us,Tamanini:2012hg}. It has been shown that suitable teleparallel geometries for this case can be obtained purely from spacetime symmetries~\cite{Hohmann:2019nat}, and a full classification of all such cosmologically symmetric teleparallel geometries has been constructed~\cite{Hohmann:2020zre}. Also the cosmological dynamics for several teleparallel gravity theories based on such spatially curved geometries have been derived~\cite{Hohmann:2018rwf,Capozziello:2018hly}. However, these studies have so far been restricted to the pure background dynamics, with no treatment of perturbations.

The aim of this article is to fill this gap between the two aforementioned lines of research - perturbations of spatially flat teleparallel cosmology on one side, and unperturbed spatially curved teleparallel cosmology on the other side - and to lay the foundation of the general theory of teleparallel cosmological perturbations, by extending the perturbation theory to the spatially curved case. For this purpose we apply a transformation which has also been used in the post-Newtonian perturbation theory of teleparallel gravity~\cite{Ualikhanova:2019ygl,Emtsova:2019qsl,Flathmann:2019khc,Hohmann:2019qgo,Bahamonde:2020cfv}, by which the perturbation of the tetrad is replaced by a pure spacetime tensor, and which absorbs the otherwise complicated form of the non-diagonal tetrad. We then apply the formalism of gauge-invariant cosmological perturbations, which allows us to identify gauge-invariant quantities.

The article is structured as follows. We declare our conventions and notation for the split of spacetime and tensor fields in section~\ref{sec:split}. We then continue with a brief review of teleparallel geometry with cosmological symmetry in section~\ref{sec:telecosmo}. These two sections form the starting point for the derivation which we perform in the following sections. We define perturbations of the most general cosmologically symmetric teleparallel geometries in section~\ref{sec:pert}. In section~\ref{sec:gauge}, we show how these perturbations behave under gauge transformations, and derive gauge-invariant combinations. The application of our findings to teleparallel gravity theories is discussed in section~\ref{sec:gravity}. We end with a summary and outlook in section~\ref{sec:conclusion}.

\section{Space-time decomposition}\label{sec:split}
In order to work with perturbations around a cosmologically symmetric background geometry, a decomposition of quantities into their space and time components is commonly used. Different conventions are abundant in the literature. The aim of this section is to display the conventions and notations we use in this article. In section~\ref{ssec:stdecom}, we introduce the split of the spacetime manifold and the coordinates we use. The conventions for the cosmologically symmetric background metric are given in section~\ref{ssec:metdecom}. We then define the notations for decomposing tensor fields in section~\ref{ssec:tensdecom} and their covariant derivatives in section~\ref{ssec:derdecom}. We end with a note on the choice of the time coordinate in section~\ref{ssec:time}.

\subsection{Spacetime and coordinate decomposition}\label{ssec:stdecom}
Throughout this article we will assume a globally hyperbolic spacetime manifold \(M\), which is diffeomorphic to the product manifold \(\mathbb{R} \times \Sigma\), where \(\Sigma\) is a maximally symmetric, three-dimensional manifold. We will denote by \(i: \mathbb{R} \times \Sigma \to M\) the diffeomorphism relating these manifolds. We define the time coordinate \(t\) on \(M\) as the projection \(t = \pr_1 \circ i^{-1}: M \to \mathbb{R}\). Further, we equip \(\Sigma\) with local coordinates \((x^a)\). Together with the time coordinate, these equip \(M\) with local coordinates \((x^{\mu}) = (t, x^a)\). Note that lowercase Greek indices label coordinates on spacetime \(M\), while lowercase Latin indices label coordinates on space \(\Sigma\).

\subsection{Metric decomposition}\label{ssec:metdecom}
The space manifold \(\Sigma\) is taken to be maximally symmetric, which means that it is equipped with a metric
\begin{equation}
\gamma = \gamma_{ab}\dd x^a \otimes \dd x^b
\end{equation}
admitting the maximal number of Killing vector fields. On the spacetime manifold \(M\), we assume a Friedmann-Lema\^itre-Robertson-Walker metric \(g\) of signature \((-,+,+,+)\), whose components split as
\begin{equation}
g_{\mu\nu} = -n_{\mu}n_{\nu} + h_{\mu\nu}
\end{equation}
into the hypersurface (co-)normal
\begin{equation}
n^{\mu}\partial_{\mu} = N^{-1}\partial_t\,, \quad
n_{\mu}\dd x^{\mu} = -N\,\dd t
\end{equation}
and the spatial metric
\begin{equation}
h_{\mu\nu}\dd x^{\mu} \otimes \dd x^{\nu} = A^2\gamma_{ab}\dd x^a \otimes \dd x^b\,,
\end{equation}
where \(N = N(t)\) and \(A = A(t)\) are the lapse and the scale factor, respectively. They satisfy the normalization \(n_{\mu}n^{\mu} = -1\) and orthogonality \(n^{\mu}h_{\mu\nu} = 0\). The metric is thus given by
\begin{equation}
g_{\mu\nu}\dd x^{\mu} \otimes \dd x^{\nu} = -N^2\dd t \otimes \dd t + A^2\gamma_{ab}\dd x^a \otimes \dd x^b\,.
\end{equation}
Further, we denote by \(\epsilon_{\mu\nu\rho\sigma}\) the totally antisymmetric tensor defined by the spacetime metric \(g_{\mu\nu}\). Its spatial part is denoted \(\varepsilon_{\mu\nu\rho}\) and defined by
\begin{equation}
\varepsilon_{\mu\nu\rho} = n^{\sigma}\epsilon_{\sigma\mu\nu\rho}\,, \quad
\epsilon_{\mu\nu\rho\sigma} = 4\varepsilon_{[\mu\nu\rho}n_{\sigma]}\,.
\end{equation}
Note that, by construction, \(n^{\mu}\varepsilon_{\mu\nu\rho} = 0\). Finally, we remark that \(\varepsilon_{\mu\nu\rho}\) is related to the totally antisymmetric tensor \(\upsilon_{abc}\) of the metric \(\gamma_{ab}\) via
\begin{equation}
\varepsilon_{\mu\nu\rho}\dd x^{\mu} \otimes \dd x^{\nu} \otimes \dd x^{\rho} = A^3\upsilon_{abc}\dd x^a \otimes \dd x^b \otimes \dd x^c\,.
\end{equation}
We will make use of these relations in the following sections.

\subsection{Tensor decomposition}\label{ssec:tensdecom}
In order to conveniently relate tensor fields on the spacetime \(M\) to their decomposition into spatial and temporal components, we now introduce a prescription and corresponding notation which we will make use of in this article. This is done by first introducing the spatial tensor fields
\begin{equation}
\Pi_{\mu}^a\partial_a \otimes \dd x^{\mu} = A\delta^a_b\,\partial_a \otimes \dd x^b\,, \quad
\Pi^{\mu}_a\partial_{\mu} \otimes \dd x^a = A^{-1}\delta_a^b\,\partial_b \otimes \dd x^a\,.
\end{equation}
It follows that they are related to the unit (co-)normal \(n_{\mu}\) and induced spatial metric \(h_{\mu\nu}\) by
\begin{equation}
n_{\mu}\Pi^{\mu}_a = 0\,, \quad
n^{\mu}\Pi_{\mu}^a = 0\,, \quad
h_{\mu\nu}\Pi^{\mu}_a\Pi^{\nu}_b = \gamma_{ab}\,, \quad
\gamma_{ab}\Pi_{\mu}^a\Pi_{\nu}^b = h_{\mu\nu}\,.
\end{equation}
With their help we can write the temporal and spatial projectors in the form
\begin{equation}
\delta^{\mu}_{\nu} = -n^{\mu}n_{\nu} + h^{\mu}_{\nu} = -n^{\mu}n_{\nu} + \Pi^{\mu}_a\Pi_{\nu}^a\,, \quad
\Pi^a_{\mu}\Pi_b^{\mu} = \delta^a_b\,.
\end{equation}
For a vector field \(X = X^{\mu}\partial_{\mu}\) we then introduce the notation
\begin{equation}
X = N^{-1}\hat{X}^0\partial_t + A^{-1}\hat{X}^a\partial_a\,, \quad
\hat{X}^0 = -n_{\mu}X^{\mu} = NX^0\,, \quad
\hat{X}^a = \Pi_{\mu}^aX^{\mu} = AX^a
\end{equation}
for the temporal and spatial components. Conversely, for a covector field \(\alpha = \alpha_{\mu}\dd x^{\mu}\) we write
\begin{equation}
\alpha = N\hat{\alpha}_0\,\dd t + A\hat{\alpha}_a\,\dd x^a\,, \quad
\hat{\alpha}_0 = n^{\mu}\alpha_{\mu} = N^{-1}\alpha_0\,, \quad
\hat{\alpha}_a = \Pi^{\mu}_a\alpha_{\mu} = A^{-1}\alpha_a\,.
\end{equation}
The advantage of these definitions becomes apparent when it comes to raising and lowering indices using the metric tensor. Using the so-called musical isomorphisms, which are defined by
\begin{subequations}
\begin{align}
X^{\flat} &= X_{\mu}\,\dd x^{\mu} = g_{\mu\nu}X^{\mu}\,\dd x^{\nu} = -N^2X^0\,\dd t + A^2\gamma_{ab}X^a\,\dd x^b\,,\\
\alpha^{\sharp} &= \alpha^{\mu}\partial_{\mu} = g^{\mu\nu}\alpha_{\mu}\partial_{\nu} = -N^{-2}\alpha_0\partial_t + A^{-2}\gamma^{ab}\alpha_a\partial_b\,,
\end{align}
\end{subequations}
we see that the temporal and spatial components are related by
\begin{equation}
\hat{X}^0 = -\hat{X}_0\,, \quad
\hat{X}^a = \gamma^{ab}\hat{X}_b\,,
\end{equation}
and analogously for \(\alpha\). Hence, for the decomposed tensor fields indices are raised and lowered with the background metric \(-\dd t \otimes \dd t + \gamma_{ab}\dd x^a \otimes \dd x^b\). Note that the components of the latter are independent of the time coordinate \(t\), in contrast to the components of the spacetime metric \(g_{\mu\nu}\). Raising and lowering the indices with the background metric therefore has the advantage that it commutes with the time derivative \(\partial_t\). We will make use of this fact in the remainder of this article, and in particular in the decomposition of derivatives.

\subsection{Derivative decomposition}\label{ssec:derdecom}
We now make use of the tensor decomposition introduced above to decompose their covariant derivatives. On the spacetime manifold \(M\), we denote the Levi-Civita covariant derivative of the metric \(g_{\mu\nu}\) by \(\lc{\nabla}_{\mu}\), so that\footnote{We use the convention that the last lower index on any connection coefficient \(\lc{\Gamma}^{\mu}{}_{\nu\rho}\) is the ``derivative'' index (which does not make a difference for the Levi-Civita connection, since it is symmetric, but will be relevant for the teleparallel connection used later).}
\begin{equation}
\lc{\nabla}_{\mu}X^{\nu} = \partial_{\mu}X^{\nu} + \lc{\Gamma}^{\nu}{}_{\rho\mu}X^{\rho}
\end{equation}
for vector fields \(X^{\mu}\) on \(M\), with
\begin{equation}
\lc{\Gamma}^{\mu}{}_{\nu\rho} = \frac{1}{2}g^{\mu\sigma}(\partial_{\nu}g_{\sigma\rho} + \partial_{\rho}g_{\nu\sigma} - \partial_{\sigma}g_{\nu\rho})\,.
\end{equation}
Given two vector fields \(X^{\mu}\) and \(Y^{\mu}\) on \(M\), which are spatial, \(n_{\mu}X^{\mu} = n_{\mu}Y^{\mu} = 0\), the covariant derivative
\begin{equation}
\lc{\nabla}_XY = (X^{\mu}\lc{\nabla}_{\mu}Y^{\nu})\partial_{\nu}
\end{equation}
is not necessarily spatial. Its decomposition
\begin{equation}
\lc{\nabla}_XY = \lc{\DD}_XY + nK(X, Y)
\end{equation}
defines the spatial covariant derivative
\begin{equation}
\lc{\DD}_XY = (X^{\mu}\lc{\DD}_{\mu}Y^{\nu})\partial_{\nu} = (X^{\mu}\lc{\nabla}_{\mu}Y^{\nu})h_{\nu}^{\rho}\partial_{\rho}\,,
\end{equation}
as well as the intrinsic curvature
\begin{equation}
K(X, Y) = K_{\mu\nu}X^{\mu}Y^{\nu} = -n_{\nu}X^{\mu}\lc{\nabla}_{\mu}Y^{\nu}\,.
\end{equation}
The latter is related to the acceleration vector field \(a_{\mu}\) by
\begin{equation}
K_{\mu\nu} = \lc{\nabla}_{\mu}n_{\nu} + n_{\mu}a_{\nu}\,, \quad
a_{\mu} = n^{\nu}\lc{\nabla}_{\nu}n_{\mu}\,.
\end{equation}
In the case of the Friedmann-Lema\^itre-Robertson-Walker metric we consider here, the acceleration and extrinsic curvature are given by
\begin{equation}
a_{\mu} = 0\,, \quad
K_{\mu\nu} = Hh_{\mu\nu}\,, \quad
H = \frac{\partial_tA}{NA}\,,
\end{equation}
where \(H\) is the Hubble parameter.

In order to relate the introduced decomposition of covariant derivatives on spacetime \(M\) to those projected on the spatial manifold \(\Sigma\), we denote by \(\dd_a\) the Levi-Civita derivative of the metric \(\gamma_{ab}\), which acts on vector fields \(\xi^a\) on \(\Sigma\) as
\begin{equation}
\dd_a\xi^b = \partial_a\xi^b + \Gamma^b{}_{ca}\xi^c\,,
\end{equation}
with the Christoffel symbols
\begin{equation}
\Gamma^a{}_{bc} = \frac{1}{2}\gamma^{ad}(\partial_b\gamma_{dc} + \partial_c\gamma_{bd} - \partial_d\gamma_{bc})\,.
\end{equation}
For a vector field \(X^{\mu}\) we then find the helpful relation
\begin{equation}\label{eq:derdecom}
\Pi_a^{\mu}\Pi^b_{\nu}\lc{\nabla}_{\mu}(h^{\nu}_{\rho}X^{\rho}) = A^{-1}\dd_a(\Pi^b_{\mu}X^{\mu}) = A^{-1}\dd_a\hat{X}^b\,,
\end{equation}
which we will use later.

\subsection{Choice of the time coordinate}\label{ssec:time}
Throughout this article, we will be using an arbitrary time coordinate \(t\), which is signaled by the presence of the arbitrary lapse function \(N\). While this choice leads to the necessity of explicitly carrying the lapse function through all calculations, it has the advantage that it will simplify the comparison of our results with the literature, where essentially two different choices of the time coordinate are present. These are the cosmological time \(\hat{t}\) and the conformal time \(\mathfrak{t}\), which are related to the coordinate time by
\begin{equation}
\dd\hat{t} = N\,\dd t = A\,\dd\mathfrak{t}\,.
\end{equation}
Hence, working in the cosmological time, \(t \equiv \hat{t}\), fixes the lapse function to \(N \equiv 1\), while in the conformal time \(t \equiv \mathfrak{t}\) one has \(N \equiv A\). It is conventional to denote derivatives with respect to coordinate time by a dot and by conformal time by a prime, hence for a scalar function \(f = f(t)\) one has
\begin{equation}\label{eq:timeder}
\dot{f} = \frac{\dd f}{\dd\hat{t}} = \frac{1}{N}\partial_tf = \mathcal{L}_nf\,, \quad
f' = \frac{\dd f}{\dd\mathfrak{t}} = \frac{A}{N}\partial_tf\,.
\end{equation}
Note that due to the presence of time-dependent numerical factors, higher order derivatives will incur derivatives of these factors. In particular, we will also define the conformal Hubble parameter by
\begin{equation}\label{eq:chubble}
\mathcal{H} = \frac{A'}{A} = \dot{A} = AH\,.
\end{equation}
In the following sections, we will make use of the conventions laid out in this section and apply them to the teleparallel geometry.

\section{Teleparallel geometry and cosmology}\label{sec:telecosmo}
Before we study perturbations of cosmologically symmetric teleparallel geometry in the next section, we now briefly review the background geometry. The definitions and conventions used for the teleparallel geometry are summarized in section~\ref{ssec:tele}. We then briefly discuss cosmological symmetry in the context of teleparallel geometry in section~\ref{ssec:cosmo}. The particular cosmologically symmetric geometries whose perturbations we will study are summarized in section~\ref{ssec:symtet}.

\subsection{Teleparallel geometry}\label{ssec:tele}
We start with a brief overview of the essential ingredients of teleparallel geometry; see~\cite{Aldrovandi:2013wha} for a comprehensive introduction. In the covariant formulation of teleparallel gravity~\cite{Krssak:2015oua,Hohmann:2018rwf,Krssak:2018ywd}, one assumes that the fundamental, dynamical fields which describe the geometry of the spacetime \(M\) are the tetrad \(\theta^A = \theta^A{}_{\mu}\dd x^{\mu}\) and the teleparallel spin connection~\(\tp{\omega}^A{}_B = \tp{\omega}^A{}_{B\mu}\dd x^{\mu}\), where capital Latin letters denote Lorentz indices \(0, \ldots, 3\). Together with the Minkowski metric \(\eta_{AB} = \mathrm{diag}(-1,1,1,1)\) and the inverse tetrad \(e_A = e_A{}^{\mu}\partial_{\mu}\), which satisfies
\begin{equation}
e_A \intprod \theta^B = e_A{}^{\mu}\theta^B{}_{\mu} = \delta_A^B\,,
\end{equation}
these define the metric
\begin{equation}\label{eq:metric}
g_{\mu\nu} = \eta_{AB}\theta^A{}_{\mu}\theta^B{}_{\nu}
\end{equation}
and the coefficients
\begin{equation}\label{eq:affconn}
\tp{\Gamma}^{\mu}{}_{\nu\rho} = e_A{}^{\mu}\left(\partial_{\rho}\theta^A{}_{\nu} + \tp{\omega}^A{}_{B\rho}\theta^B{}_{\nu}\right)
\end{equation}
of an affine connection. The spin connection is assumed to be metric compatible,
\begin{equation}\label{eq:nmform}
\tp{Q}_{AB} = \eta_{CB}\tp{\omega}^C{}_A + \eta_{AC}\tp{\omega}^C{}_{B} \equiv 0\,,
\end{equation}
and flat
\begin{equation}\label{eq:curvform}
\tp{R}^A{}_B = \dd\tp{\omega}^A{}_B + \tp{\omega}^A{}_C \wedge \tp{\omega}^C{}_B \equiv 0\,.
\end{equation}
This implies that the same properties hold for the affine connection. Hence,
\begin{equation}\label{eq:affnonmet}
\tp{Q}_{\rho\mu\nu} = \tp{\nabla}_{\rho}g_{\mu\nu} \equiv 0\,,
\end{equation}
and
\begin{equation}\label{eq:affcurv}
\tp{R}^{\rho}{}_{\sigma\mu\nu} = \partial_{\mu}\tp{\Gamma}^{\rho}{}_{\sigma\nu} - \partial_{\nu}\tp{\Gamma}^{\rho}{}_{\sigma\mu} + \tp{\Gamma}^{\rho}{}_{\tau\mu}\tp{\Gamma}^{\tau}{}_{\sigma\nu} - \tp{\Gamma}^{\rho}{}_{\tau\nu}\tp{\Gamma}^{\tau}{}_{\sigma\mu} \equiv 0\,.
\end{equation}
It does have, however, non-vanishing torsion, which is expressed in differential forms as
\begin{equation}\label{eq:torsform}
\tp{T}^A = \dd\theta^A + \tp{\omega}^A{}_B \wedge \theta^B\,,
\end{equation}
or equivalently through the affine connection as
\begin{equation}\label{eq:torsion}
\tp{T}^{\rho}{}_{\mu\nu} = \tp{\Gamma}^{\rho}{}_{\nu\mu} - \tp{\Gamma}^{\rho}{}_{\mu\nu}\,.
\end{equation}
For later use we further define the contortion tensor
\begin{equation}\label{eq:contor}
\tp{K}^{\mu}{}_{\nu\rho} = \tp{\Gamma}^{\mu}{}_{\nu\rho} - \lc{\Gamma}^{\mu}{}_{\nu\rho} = \frac{1}{2}\left(\tp{T}_{\nu}{}^{\mu}{}_{\rho} + \tp{T}_{\rho}{}^{\mu}{}_{\nu} - \tp{T}^{\mu}{}_{\nu\rho}\right)\,,
\end{equation}
which measures the difference between the teleparallel and Levi-Civita connections.

An important property of teleparallel gravity theories in the covariant formulation is their invariance under local Lorentz transformations \(\Lambda^A{}_B\), which act on the tetrad and spin connection via
\begin{equation}
\theta'^A{}_{\mu} = \Lambda^A{}_B\theta^B{}_{\mu}\,, \quad
\tp{\omega}'^A{}_{B\mu} = \Lambda^A{}_C(\Lambda^{-1})^D{}_B\tp{\omega}^C{}_{D\mu} + \Lambda^A{}_C\partial_{\mu}(\Lambda^{-1})^C{}_B\,.
\end{equation}
Together with the metricity and flatness of the spin connection this allows to find a Lorentz transformation such that the spin connection vanishes. In the remainder of this article, we will make this choice and impose the Weitzenböck gauge condition \(\tp{\omega}^A{}_{B\mu} \equiv 0\) both at the background and perturbation levels; see~\cite{Golovnev:2018wbh} for a discussion of the role of the spin connection for cosmological perturbations in teleparallel gravity.

\subsection{Cosmological symmetry}\label{ssec:cosmo}
Following the treatment in~\cite{Hohmann:2015pva,Hohmann:2019nat}, infinitesimal symmetries of teleparallel geometries are defined as vector fields \(X^{\mu}\) on \(M\) which leave the metric and the affine connection invariant,
\begin{equation}\label{eq:symmet}
0 = (\mathcal{L}_Xg)_{\mu\nu} = X^{\rho}\partial_{\rho}g_{\mu\nu} + \partial_{\mu}X^{\rho}g_{\rho\nu} + \partial_{\nu}X^{\rho}g_{\mu\rho}
\end{equation}
and
\begin{equation}\label{eq:symconn}
0 = (\mathcal{L}_X\tp{\Gamma})^{\mu}{}_{\nu\rho} = X^{\sigma}\partial_{\sigma}\tp{\Gamma}^{\mu}{}_{\nu\rho} - \partial_{\sigma}X^{\mu}\tp{\Gamma}^{\sigma}{}_{\nu\rho} + \partial_{\nu}X^{\sigma}\tp{\Gamma}^{\mu}{}_{\sigma\rho} + \partial_{\rho}X^{\sigma}\tp{\Gamma}^{\mu}{}_{\nu\sigma} + \partial_{\nu}\partial_{\rho}X^{\mu}.
\end{equation}
While the former is simply the well-known statement that \(X\) is a Killing vector field, the latter is less common. It turns out to be a tensor field~\cite{Yano:1957lda}, so that the demand that is vanishes is indeed meaningful independent of the choice of coordinates, in contrast to connection coefficients.

For the cosmological symmetry we study here, we can express the generating vector fields most easily by choosing the spherical spatial coordinates \((x^a) = (r, \vartheta, \varphi)\). In these coordinates, the generators of rotations read
\begin{subequations}\label{eq:genrot}
\begin{align}
R_1 &= \sin\varphi\partial_{\vartheta} + \frac{\cos\varphi}{\tan\vartheta}\partial_{\varphi}\,,\\
R_2 &= -\cos\varphi\partial_{\vartheta} + \frac{\sin\varphi}{\tan\vartheta}\partial_{\varphi}\,,\\
R_3 &= -\partial_{\varphi}\,,
\end{align}
\end{subequations}
while the generators of translations are given by
\begin{subequations}\label{eq:gentra}
\begin{align}
T_1 &= \chi\sin\vartheta\cos\varphi\partial_r + \frac{\chi}{r}\cos\vartheta\cos\varphi\partial_{\vartheta} - \frac{\chi\sin\varphi}{r\sin\vartheta}\partial_{\varphi}\,,\\
T_2 &= \chi\sin\vartheta\sin\varphi\partial_r + \frac{\chi}{r}\cos\vartheta\sin\varphi\partial_{\vartheta} + \frac{\chi\cos\varphi}{r\sin\vartheta}\partial_{\varphi}\,,\\
T_3 &= \chi\cos\vartheta\partial_r - \frac{\chi}{r}\sin\vartheta\partial_{\vartheta}\,,
\end{align}
\end{subequations}
where \(\chi = \sqrt{1 - u^2r^2}\) and \(u\) is an arbitrary real or imaginary constant. More commonly, this parameter is denoted \(k = u^2 \in \{-1, 0, 1\}\), where the three possible values correspond to the sign of the curvature of the maximally symmetric metric \(\gamma_{ab}\) on the space \(\Sigma\).

Imposing the symmetry conditions~\eqref{eq:symmet} and~\eqref{eq:symconn} for the symmetry generators~\eqref{eq:genrot} and~\eqref{eq:gentra} significantly restricts the allowed possible teleparallel geometries. A complete list of all possible solutions has been derived in~\cite{Hohmann:2020zre}. They have in common that their torsion tensors, and consequently their contortion tensors, are of the form
\begin{equation}\label{eq:torcosmo}
\tp{\bar{T}}_{\mu\nu\rho} = \frac{2\mathscr{V}h_{\mu[\nu}n_{\rho]} + 2\mathscr{A}\varepsilon_{\mu\nu\rho}}{A}\,, \quad
\tp{\bar{K}}_{\mu\nu\rho} = \frac{2\mathscr{V}h_{\rho[\mu}n_{\nu]} - \mathscr{A}\varepsilon_{\mu\nu\rho}}{A}\,,
\end{equation}
where we use a bar to denote quantities representing the cosmologically symmetric background, with functions \(\mathscr{V} = \mathscr{V}(t)\) and \(\mathscr{A} = \mathscr{A}(t)\), which determine the vector and axial parts of the torsion, and which are further restricted by the condition that the teleparallel connection must be flat. The particular choice of the numerical factors will become clear in the following sections, where we will list the values of these functions for all cosmologically symmetric teleparallel geometries, and when they will enter the equations which govern the evolution of perturbations.

\subsection{Cosmologically symmetric tetrads}\label{ssec:symtet}
For reference, we finally list the cosmologically symmetric tetrads whose perturbations we consider, together with the implied properties of the functions \(\mathscr{V}\) and \(\mathscr{A}\) introduced above; see~\cite{Hohmann:2020zre} for a comprehensive derivation of these tetrads and their properties:
\begin{enumerate}
\item
The ``vector'' tetrad is given by
\begin{subequations}\label{eq:tetwbvec}
\begin{align}
\bar{\theta}^0 &= N\chi\dd t + iuA\frac{r}{\chi}\dd r\,,\\
\bar{\theta}^1 &= A\left[\sin\vartheta\cos\varphi\left(\dd r + iu\frac{N}{A}r\dd t\right) + r\cos\vartheta\cos\varphi\dd\vartheta - r\sin\vartheta\sin\varphi\dd\varphi\right]\,,\\
\bar{\theta}^2 &= A\left[\sin\vartheta\sin\varphi\left(\dd r + iu\frac{N}{A}r\dd t\right) + r\cos\vartheta\sin\varphi\dd\vartheta + r\sin\vartheta\cos\varphi\dd\varphi\right]\,,\\
\bar{\theta}^3 &= A\left[\cos\vartheta\left(\dd r + iu\frac{N}{A}r\dd t\right) - r\sin\vartheta\dd\vartheta\right]\,.
\end{align}
\end{subequations}
It is characterized by having only vector torsion,
\begin{equation}\label{eq:tetfuncvec}
\mathscr{V} = \mathcal{H} \pm iu\,, \quad
\mathscr{A} = 0\,,
\end{equation}
where \(\mathcal{H}\) is the conformal Hubble parameter~\eqref{eq:chubble}. Note that this tetrad, and also its torsion, is real if and only if \(u\) is imaginary, hence if the spatial curvature \(k = u^2 \leq 0\) is non-positive.

\item
The ``axial'' tetrad is given by
\begin{subequations}\label{eq:tetwbaxi}
\begin{align}
\bar{\theta}^0 &= N\dd t\,,\\
\bar{\theta}^1 &= A\left[\frac{\sin\vartheta\cos\varphi}{\chi}\dd r + r(\chi\cos\vartheta\cos\varphi + ur\sin\varphi)\dd\vartheta - r\sin\vartheta(\chi\sin\varphi - ur\cos\vartheta\cos\varphi)\dd\varphi\right]\,,\\
\bar{\theta}^2 &= A\left[\frac{\sin\vartheta\sin\varphi}{\chi}\dd r + r(\chi\cos\vartheta\sin\varphi - ur\cos\varphi)\dd\vartheta + r\sin\vartheta(\chi\cos\varphi + ur\cos\vartheta\sin\varphi)\dd\varphi\right]\,,\\
\bar{\theta}^3 &= A\left[\frac{\cos\vartheta}{\chi}\dd r - r\chi\sin\vartheta\dd\vartheta - ur^2\sin^2\vartheta\dd\varphi\right]\,.
\end{align}
\end{subequations}
This tetrad also has axial torsion, in addition to the vector torsion, where
\begin{equation}\label{eq:tetfuncaxi}
\mathscr{V} = \mathcal{H}\,, \quad
\mathscr{A} = \pm u\,.
\end{equation}
In contrast to the previous branch, it is real if and only if \(u\) is real, hence for non-negative spatial curvature \(k = u^2 \geq 0\).

\item
We finally mention that in the limit \(u = 0\) the two solution branches have a common intersection
\begin{subequations}\label{eq:tetwbflat}
\begin{align}
\bar{\theta}^0 &= N\dd t\,,\\
\bar{\theta}^1 &= A\left[\sin\vartheta\cos\varphi\dd r + r\cos\vartheta\cos\varphi\dd\vartheta - r\sin\vartheta\sin\varphi\dd\varphi\right]\,,\\
\bar{\theta}^2 &= A\left[\sin\vartheta\sin\varphi\dd r + r\cos\vartheta\sin\varphi\dd\vartheta + r\sin\vartheta\cos\varphi\dd\varphi\right]\,,\\
\bar{\theta}^3 &= A\left[\cos\vartheta\dd r - r\sin\vartheta\dd\vartheta\right]\,,
\end{align}
whose irreducible torsion components are given by
\end{subequations}
\begin{equation}\label{eq:tetfuncflat}
\mathscr{V} = \mathcal{H}\,, \quad
\mathscr{A} = 0\,.
\end{equation}
This flat limiting case is equivalent to the flat, diagonal tetrad in Cartesian coordinates~\cite{Hohmann:2019nat}.
\end{enumerate}
This concludes our review of the cosmologically symmetric teleparallel geometry. With these definitions and conventions in place, we can now come to perturbations around the cosmologically symmetric background in the following section.

\section{Cosmological tetrad perturbations}\label{sec:pert}
We now come to perturbations of the cosmologically symmetric teleparallel geometry displayed in the previous section. Since we work in the Weitzenböck gauge at all perturbation orders, it will be sufficient to discuss perturbations of the tetrad. We give their definition in section~\ref{ssec:perdef}. The expansion of several relevant tensor fields, which appear as constituents of the field equations of teleparallel gravity theories, is shown in section~\ref{ssec:pertele}, showing that these become independent of the particular choice of the Lorentz frame, as they are expressed purely in spacetime components. We then perform an irreducible decomposition of the perturbations in section~\ref{ssec:perdec}.

\subsection{Definition}\label{ssec:perdef}
The most direct approach towards cosmological perturbations in teleparallel geometry would be to define them as perturbations of the tetrad components around the cosmological background given in section~\ref{ssec:symtet}, in analogy to the standard practice for metric perturbations. While this is straightforward in the case of a flat, diagonal background tetrad~\cite{Golovnev:2018wbh,Golovnev:2020aon}, it would be rather cumbersome for the general, non-diagonal background tetrads. Further, it would depend on a particular choice of coordinates. Hence, we propose a different approach here. For this purpose we proceed in analogy to the post-Newtonian tetrad perturbation approach~\cite{Ualikhanova:2019ygl,Emtsova:2019qsl,Flathmann:2019khc,Hohmann:2019qgo,Bahamonde:2020cfv} and first define the linear tetrad perturbation as
\begin{equation}\label{eq:tetpert}
\theta^A{}_{\mu} = \bar{\theta}^A{}_{\mu} + \delta\theta^A{}_{\mu} = \bar{\theta}^A{}_{\mu} + \tau^A{}_{\mu}\,,
\end{equation}
around the cosmologically symmetric background tetrad \(\bar{\theta}^A{}_{\mu}\). Only terms linear in the perturbations \(\tau^A{}_{\mu}\) will be further considered. It is then helpful to lower the Lorentz index using the Minkowski metric \(\eta_{AB}\) and transform it to a spacetime index using the background tetrad \(\bar{\theta}^A{}_{\mu}\), in order to obtain
\begin{equation}\label{eq:perttens}
\tau_{\mu\nu} = \eta_{AB}\bar{\theta}^A{}_{\mu}\tau^B{}_{\nu}\,.
\end{equation}
In the following, we will expand all relevant quantities in terms of \(\tau_{\mu\nu}\).

\subsection{Perturbation of the teleparallel geometry}\label{ssec:pertele}
It is instructive to derive the linear perturbations of the geometric objects introduced in section~\ref{ssec:tele} in terms of the tetrad perturbation~\eqref{eq:perttens} and the cosmologically symmetric background geometry shown in section~\ref{ssec:cosmo}. For the metric, one immediately finds that the perturbation is simply the symmetric part of the tetrad perturbation,
\begin{equation}\label{eq:metricpert}
\delta g_{\mu\nu} = 2\tau_{(\mu\nu)}\,.
\end{equation}
This can be used for calculating the perturbation of the Levi-Civita connection. The latter can most easily be expressed as the covariant derivative of the metric perturbation, and thus reads
\begin{equation}
\delta\lc{\Gamma}^{\rho}{}_{\mu\nu} = \frac{1}{2}\bar{g}^{\rho\sigma}\left(\lc{\bar{\nabla}}_{\mu}\delta g_{\sigma\nu} + \lc{\bar{\nabla}}_{\nu}\delta g_{\mu\sigma} - \lc{\bar{\nabla}}_{\sigma}\delta g_{\mu\nu}\right) = \bar{g}^{\rho\sigma}\left(\lc{\bar{\nabla}}_{\mu}\tau_{(\sigma\nu)} + \lc{\bar{\nabla}}_{\nu}\tau_{(\mu\sigma)} - \lc{\bar{\nabla}}_{\sigma}\tau_{(\mu\nu)}\right)\,.
\end{equation}
For the Weitzenböck connection we find the formula
\begin{equation}\label{eq:tpconpert}
\delta\tp{\Gamma}^{\mu}{}_{\nu\rho} = \tp{\bar{\nabla}}_{\rho}\tau^{\mu}{}_{\nu} = \lc{\bar{\nabla}}_{\rho}\tau^{\mu}{}_{\nu} + \tp{\bar{K}}^{\mu}{}_{\sigma\rho}\tau^{\sigma}{}_{\nu} - \tp{\bar{K}}^{\sigma}{}_{\nu\rho}\tau^{\mu}{}_{\sigma}\,.
\end{equation}
Taking the part which is antisymmetric in the lower two indices, we obtain the torsion perturbation
\begin{equation}\label{eq:torspert}
\delta\tp{T}^{\mu}{}_{\nu\rho} = 2\tp{\bar{\nabla}}_{[\nu}\tau^{\mu}{}_{\rho]} = 2\lc{\bar{\nabla}}_{[\nu}\tau^{\mu}{}_{\rho]} + 2\tp{\bar{K}}^{\mu}{}_{\sigma[\nu}\tau^{\sigma}{}_{\rho]} - 2\tp{\bar{K}}^{\sigma}{}_{[\rho\nu]}\tau^{\mu}{}_{\sigma}\,.
\end{equation}
Finally, we also display the perturbation of the covariant derivative of the torsion, as it will enter into the gravitational field equations discussed in section~\ref{sec:gravity}. Note that here both the perturbation of the torsion tensor and of the connection coefficients appearing in the covariant derivative must be taken into account. This is done most easily using the teleparallel connection. Using the formulas~\eqref{eq:tpconpert} and~\eqref{eq:torspert} one finds
\begin{equation}
\begin{split}
\delta\tp{\nabla}_{\sigma}\tp{T}^{\mu}{}_{\nu\rho} &= \tp{\bar{\nabla}}_{\sigma}\delta\tp{T}^{\mu}{}_{\nu\rho} + \delta\tp{\Gamma}^{\mu}{}_{\omega\sigma}\tp{\bar{T}}^{\omega}{}_{\nu\rho} - \delta\tp{\Gamma}^{\omega}{}_{\nu\sigma}\tp{\bar{T}}^{\mu}{}_{\omega\rho} - \delta\tp{\Gamma}^{\omega}{}_{\rho\sigma}\tp{\bar{T}}^{\mu}{}_{\nu\omega}\\
&= 2\tp{\bar{\nabla}}_{\sigma}\tp{\bar{\nabla}}_{[\nu}\tau^{\mu}{}_{\rho]} + \tp{\bar{T}}^{\omega}{}_{\nu\rho}\tp{\bar{\nabla}}_{\sigma}\tau^{\mu}{}_{\omega} - \tp{\bar{T}}^{\mu}{}_{\omega\rho}\tp{\bar{\nabla}}_{\sigma}\tau^{\omega}{}_{\nu} - \tp{\bar{T}}^{\mu}{}_{\nu\omega}\tp{\bar{\nabla}}_{\sigma}\tau^{\omega}{}_{\rho}\,.
\end{split}
\end{equation}
Note that all perturbations are expressed in terms of the perturbation~\eqref{eq:perttens}, which carries only spacetime indices and no Lorentz indices, and background objects which obey the cosmological symmetry. It is a consequence of this fact that the cumbersome explicit coordinate expressions of the tetrad shown in section~\ref{ssec:symtet} will not enter the further calculations, and the properties of the cosmological background are fully modeled by the functions \(N, A, \mathscr{V}, \mathscr{A}\) which characterize the background geometry.

\subsection{Irreducible decomposition}\label{ssec:perdec}
A crucial step in the theory of cosmological perturbations is their decomposition into components which form irreducible representations of the rotation group, acting on the spatial hypersurfaces. We now perform this decomposition for the tetrad perturbations~\eqref{eq:perttens}, following a similar scheme as conducted for the spatially flat background tetrad~\cite{Golovnev:2018wbh}. Using the notations introduced in section~\ref{sec:split}, we write the perturbations as
\begin{equation}
\tau_{\mu\nu} = \phi n_{\mu}n_{\nu} - (b_{\nu} + A\lc{\DD}_{\nu}j)n_{\mu} - (v_{\mu} + A\lc{\DD}_{\mu}y)n_{\nu} + \psi h_{\mu\nu} + A^2\lc{\DD}_{\mu}\lc{\DD}_{\nu}\sigma + A\lc{\DD}_{\nu}c_{\mu} + \varepsilon_{\mu\nu\rho}(A\lc{\DD}^{\rho}\xi + w^{\rho}) + \frac{1}{2}q_{\mu\nu}\,,
\end{equation}
where the vectors \(b_{\mu}, v_{\mu}, c_{\mu}, w_{\mu}\) and the symmetric, trace-free tensor \(q_{\mu\nu}\) are spatial and divergence-free,
\begin{equation}
\lc{\DD}_{\mu}b^{\mu} = \lc{\DD}_{\mu}v^{\mu} = \lc{\DD}_{\mu}c^{\mu} = \lc{\DD}_{\mu}w^{\mu} = 0\,, \quad
\lc{\DD}_{\mu}q^{\mu\nu} = 0\,, \quad
q_{[\mu\nu]} = 0\,, \quad
q_{\mu}{}^{\mu} = 0\,.
\end{equation}
Note the appearance of scale factors in front of the spatial derivatives. These are introduced such that after performing the space-time split of the perturbations as detailed in section~\ref{ssec:tensdecom}, they cancel the corresponding factors arising from the derivative decomposition~\eqref{eq:derdecom}, so that the resulting relations take the simpler form
\begin{equation}\label{eq:pertdec}
\hat{\tau}_{00} = \hat{\phi}\,, \quad
\hat{\tau}_{0b} = \dd_b\hat{j} + \hat{b}_b\,, \quad
\hat{\tau}_{a0} = \dd_a\hat{y} + \hat{v}_a\,, \quad
\hat{\tau}_{ab} = \hat{\psi}\gamma_{ab} + \dd_a\dd_b\sigma + \dd_b\hat{c}_a + \upsilon_{abc}(\dd^c\hat{\xi} + \hat{w}^c) + \frac{1}{2}\hat{q}_{ab}\,.
\end{equation}
We also remark that we do not symmetrize the contribution \(\dd_b\hat{c}_a\). Note that this does not pose any restriction on the generality of the perturbations, since any antisymmetric contribution of the form \(\dd_{[a}\hat{c}_{b]}\) can be rewritten as
\begin{equation}
\dd_{[a}\hat{c}_{b]} = \frac{1}{2}\upsilon_{abc}\upsilon^{dec}\dd_d\hat{c}_e
\end{equation}
and thus absorbed into a redefinition of \(\hat{w}_a\). Here the divergence of the last term vanishes due to the Bianchi identity
\begin{equation}
\dd_c(\upsilon^{dec}\dd_d\hat{c}_e) = \upsilon^{dec}\dd_{[c}\dd_{d]}\hat{c}_e = \frac{1}{2}\upsilon^{dec}R^f{}_{ecd}\hat{c}_f = 0\,.
\end{equation}
The reason for the particular decomposition~\eqref{eq:pertdec} will become clear in the following section, when we discuss gauge transformations.

\section{Gauge transformation and gauge invariance}\label{sec:gauge}
An important simplification of cosmological perturbation theory is achieved by decomposing the perturbations introduced in the previous section into gauge-invariant components. For this purpose, we display how they transform under infinitesimal coordinate changes in section~\ref{ssec:gaugedef}. This transformation is then decomposed into irreducible components in section~\ref{ssec:gaugedec}. Finally, we construct the gauge-invariant combinations in section~\ref{ssec:gaugeinv}.

\subsection{Definition}\label{ssec:gaugedef}
The starting point of our discussion in this section is the observation that the teleparallel geometry we study retains its form as a small perturbation of a cosmologically symmetric geometry under an infinitesimal coordinate transformation of the form
\begin{equation}\label{eq:coordtrans}
x'^{\mu} = x^{\mu} + X^{\mu}(x)\,,
\end{equation}
provided that the components of the vector field \(X^{\mu}\) are sufficiently small to preserve the order of the tetrad perturbations. Under this transformation the tetrad changes by the Lie derivative
\begin{equation}
\theta^A{}_{\mu} = \theta'^A{}_{\mu} + (\mathcal{L}_X\bar{\theta})^A{}_{\mu}\,,
\end{equation}
where only terms of at most linear order in the tetrad perturbation and the vector field \(X^{\mu}\) have been considered; hence, the tetrad has been replaced by the background tetrad \(\bar{\theta}^A{}_{\mu}\) in the second term. Writing the transformed tetrad as a perturbation of the same background tetrad,
\begin{equation}
\theta'^A{}_{\mu} = \bar{\theta}^A{}_{\mu} + \delta\theta'^A{}_{\mu} = \bar{\theta}^A{}_{\mu} + \tau'^A{}_{\mu}\,,
\end{equation}
we find that the transformation of the perturbation is given by
\begin{equation}\label{eq:pertlieder}
\delta_X\tau^A{}_{\mu} = \tau^A{}_{\mu} - \tau'^A{}_{\mu} = (\mathcal{L}_X\bar{\theta})^A{}_{\mu} = X^{\nu}\partial_{\nu}\bar{\theta}^A{}_{\mu} + \partial_{\mu}X^{\nu}\bar{\theta}^A{}_{\nu}\,.
\end{equation}
Lowering and transforming the Lorentz index with the background geometry, and replacing the partial derivatives acting on the background tetrad and the vector field by coefficients of the teleparallel affine connection and covariant derivatives, we obtain
\begin{equation}\label{eq:gaugetrans}
\delta_X\tau_{\mu\nu} = \tp{\bar{\nabla}}_{\nu}X_{\mu} - \bar{T}_{\mu\nu}{}^{\rho}X_{\rho} = \lc{\bar{\nabla}}_{\nu}X_{\mu} + \bar{K}_{\mu\nu}{}^{\rho}X_{\rho}\,.
\end{equation}
Note in particular that the symmetric and antisymmetric part transform as
\begin{equation}
\delta_X\tau_{(\mu\nu)} = \lc{\bar{\nabla}}_{(\mu}X_{\nu)} = \frac{1}{2}(\mathcal{L}_X\bar{g})_{\mu\nu}\,, \quad
\delta_X\tau_{[\mu\nu]} = \bar{K}_{\mu\nu}{}^{\rho}X_{\rho} - \lc{\bar{\nabla}}_{[\mu}X_{\nu]}\,.
\end{equation}
The former is a direct consequence of the relation~\eqref{eq:metricpert} between the perturbation \(\delta g_{\mu\nu}\) of the metric and the symmetric part of the tensor perturbation. Since it depends only on the metric background, it is independent of the choice of the different tetrad branches shown in section~\ref{ssec:symtet}. These enter only the transformation of the antisymmetric part via the contortion tensor~\eqref{eq:torcosmo}, which depends on the functions \(\mathscr{V}\) and \(\mathscr{A}\).

\subsection{Irreducible decomposition}\label{ssec:gaugedec}
We now study the transformation of the irreducible components of the tetrad perturbations introduced in section~\ref{ssec:perdec}. For this purpose, we analogously decompose the generating vector field into the components
\begin{equation}\label{eq:gaugedec}
X_{\mu} = -X_{\perp}n_{\mu} + A\lc{\DD}_{\mu}X_{\parallel} + Z_{\mu}\,,
\end{equation}
where \(X_{\perp}\) and \(X_{\parallel}\) are scalars, while \(Z_{\mu}\) is a spatial and divergence-free vector. Hence, we have
\begin{equation}
\hat{X}_0 = \hat{X}_{\perp}\,, \quad
\hat{X}_a = \dd_a\hat{X}_{\parallel} + \hat{Z}_a
\end{equation}
in the space-time decomposition. Applying the same decomposition to the gauge transformation~\eqref{eq:gaugetrans}, we find its components
\begin{gather}
\delta_X\hat{\tau}_{0b} = \frac{\dd_b\hat{X}_{\perp}}{A} + \frac{(\dd_b\hat{X}_{\parallel} + \hat{Z}_b)(N\mathscr{V} - \partial_tA)}{NA}\,, \quad
\delta_X\hat{\tau}_{a0} = \frac{\partial_t(\dd_a\hat{X}_{\parallel} + Z_a)}{N} - \frac{\mathscr{V}}{A}(\dd_a\hat{X}_{\parallel} + Z_a)\,,\nonumber\\
\delta_X\hat{\tau}_{00} = \frac{\partial_t\hat{X}_{\perp}}{N}\,, \quad
\delta_X\hat{\tau}_{ab} = \frac{\dd_b(\dd_a\hat{X}_{\parallel} + \hat{Z}_a) - \mathscr{A}\upsilon_{abc}(\dd^c\hat{X}_{\parallel} + \hat{Z}^c)}{A} - \frac{\hat{X}_{\perp}\partial_tA}{NA}\gamma_{ab}\,.
\end{gather}
Comparing these expressions with the components~\eqref{eq:pertdec} of the perturbation \(\tau_{\mu\nu}\), one finds that they obey the transformation
\begin{gather}
\delta_X\hat{\psi} = -\frac{\hat{X}_{\perp}\partial_tA}{NA}\,, \quad
\delta_X\hat{\sigma} = \frac{\hat{X}_{\parallel}}{A}\,, \quad
\delta_X\hat{y} = \frac{\partial_t\hat{X}_{\parallel}}{N} - \frac{\mathscr{V}\hat{X}_{\parallel}}{A}\,, \quad
\delta_X\hat{j} = \frac{N\hat{X}_{\perp} + (N\mathscr{V} - \partial_tA)\hat{X}_{\parallel}}{NA}\,, \quad
\delta_X\hat{\xi} = -\frac{\mathscr{A}\hat{X}_{\parallel}}{A}\,,\nonumber\\
\delta_X\hat{\phi} = \frac{\partial_t\hat{X}_{\perp}}{N}\,, \quad
\delta_X\hat{c}_a = \frac{\hat{Z}_a}{A}\,, \quad
\delta_X\hat{v}_a = \frac{\partial_t\hat{Z}_a}{N} - \frac{\mathscr{V}\hat{Z}_a}{A}\,, \quad
\delta_X\hat{b}_a = \frac{(N\mathscr{V} - \partial_tA)\hat{Z}_a}{NA}\,, \quad
\delta_X\hat{w}_a = -\frac{\mathscr{A}\hat{Z}_a}{A}\,, \quad
\delta_X\hat{q}_{ab} = 0\,.\label{eq:gtrandec}
\end{gather}
Before we study these further, it is helpful to remark that these transformations can most simply be expressed by making use of the conformal Hubble parameter~\eqref{eq:chubble} and the derivative \(\prime\) with respect to the conformal time~\eqref{eq:timeder}. In terms of these the transformation of the tetrad perturbation takes the simpler form
\begin{gather}
A\delta_X\hat{\tau}_{0b} = \dd_b\hat{X}_{\perp} + (\dd_b\hat{X}_{\parallel} + \hat{Z}_b)(\mathscr{V} - \mathcal{H})\,, \quad
A\delta_X\hat{\tau}_{a0} = \dd_a\hat{X}_{\parallel}' + Z_a' - \mathscr{V}(\dd_a\hat{X}_{\parallel} + Z_a)\,,\nonumber\\
A\delta_X\hat{\tau}_{00} = \hat{X}_{\perp}'\,, \quad
A\delta_X\hat{\tau}_{ab} = \dd_b(\dd_a\hat{X}_{\parallel} + \hat{Z}_a) - \mathscr{A}\upsilon_{abc}(\dd^c\hat{X}_{\parallel} + \hat{Z}^c) - \mathcal{H}\hat{X}_{\perp}\gamma_{ab}\,.
\end{gather}
Hence, their irreducible components are given by
\begin{gather}
A\delta_X\hat{\psi} = -\mathcal{H}\hat{X}_{\perp}\,, \quad
A\delta_X\hat{\sigma} = \hat{X}_{\parallel}\,, \quad
A\delta_X\hat{y} = \hat{X}_{\parallel}' - \mathscr{V}\hat{X}_{\parallel}\,, \quad
A\delta_X\hat{j} = \hat{X}_{\perp} + (\mathscr{V} - \mathcal{H})\hat{X}_{\parallel}\,, \quad
A\delta_X\hat{\xi} = -\mathscr{A}\hat{X}_{\parallel}\,,\nonumber\\
A\delta_X\hat{\phi} = \hat{X}_{\perp}'\,, \quad
A\delta_X\hat{c}_a = \hat{Z}_a\,, \quad
A\delta_X\hat{v}_a = \hat{Z}_a' - \mathscr{V}\hat{Z}_a\,, \quad
A\delta_X\hat{b}_a = (\mathscr{V} - \mathcal{H})\hat{Z}_a\,, \quad
A\delta_X\hat{w}_a = -\mathscr{A}\hat{Z}_a\,, \quad
A\delta_X\hat{q}_{ab} = 0\,.\label{eq:gtrandecct}
\end{gather}
A few remarks are in order. First, note the choice of including the term \(\dd_b\hat{c}_a\) without symmetrization allows for a direct identification of the transformation \(\delta_X\hat{c}_a\); this is identical to the case of a spatially flat background tetrad~\cite{Golovnev:2018wbh}. We also find a number of differences, which are signaled by the presence of the functions \(\mathscr{V}\) and \(\mathscr{A}\). We see that the component \(\hat{b}_a\), which is gauge-invariant in the spatially flat case \(\mathscr{V} = \mathcal{H}\), retains this property also for the axial branch with parameters~\eqref{eq:tetfuncaxi}, but receives a contribution from the spatial curvature parameter \(u\) in the case of the vector branch~\eqref{eq:tetfuncvec}. Similarly, one finds that for the vector branch also in the component \(\hat{j}\) a new term appears, which is absent in the other branches. Conversely, the parity-odd components \(\hat{\xi}\) and \(\hat{w}_a\), which are gauge-invariant in the spatially flat case, remain gauge-invariant only for the vector branch, but receive a non-vanishing transformation in the axial branch, which is governed by the parity-odd parameter function \(\mathscr{A}\) measuring the axial torsion of the background geometry.

\subsection{Gauge invariant perturbations}\label{ssec:gaugeinv}
From the gauge transformations of the perturbation components given above one can now easily identify gauge-invariant combinations. First, note that the tensor component \(\hat{\mathbf{q}}_{ab} = \hat{q}_{ab}\) is already gauge-invariant. We then continue with the vector components. Observe that the corresponding component \(\hat{Z}_a\) of the gauge transformation can be obtained as
\begin{equation}
\hat{Z}_a = A\delta_X\hat{c}_a\,.
\end{equation}
Inserting this relation into the transformation~\eqref{eq:gtrandec} of the remaining vector components, we find
\begin{equation}
\delta_X\hat{v}_a = \frac{A\partial_t\delta_X\hat{c}_a}{N} + \left(\frac{\partial_tA}{N} - \mathscr{V}\right)\delta_X\hat{c}_a\,,\quad
\delta_X\hat{b}_a = \left(\mathscr{V} - \frac{\partial_tA}{N}\right)\delta_X\hat{c}_a\,, \quad
\delta_X\hat{w}_a = -\mathscr{A}\delta_X\hat{c}_a\,.
\end{equation}
Hence, the combinations
\begin{equation}
\hat{\mathbf{v}}_a = \hat{v}_a + \left(\mathscr{V} - \frac{\partial_tA}{N}\right)\hat{c}_a - \frac{A}{N}\partial_t\hat{c}_a\,, \quad
\hat{\mathbf{b}}_a = \hat{b}_a + \left(\frac{\partial_tA}{N} - \mathscr{V}\right)\hat{c}_a\,, \quad
\hat{\mathbf{w}}_a = \hat{w}_a + \mathscr{A}\hat{c}_a
\end{equation}
are found to be gauge-invariant. Using the conformal time and conformal Hubble parameter, these take the form
\begin{equation}
\hat{\mathbf{v}}_a = \hat{v}_a + (\mathscr{V} - \mathcal{H})\hat{c}_a - \hat{c}_a'\,, \quad
\hat{\mathbf{b}}_a = \hat{b}_a + (\mathcal{H} - \mathscr{V})\hat{c}_a\,, \quad
\hat{\mathbf{w}}_a = \hat{w}_a + \mathscr{A}\hat{c}_a\,,
\end{equation}
which can also be seen directly from the transformations~\eqref{eq:gtrandecct}. With the scalar perturbations one can proceed similarly. The two scalar components of the gauge transformation can be expressed as
\begin{equation}
\hat{X}_{\parallel} = A\delta_X\hat{\sigma}\,, \quad
\hat{X}_{\perp} = A\left[\delta_X\hat{j} + \left(\frac{\partial_tA}{N} - \mathscr{V}\right)\delta_X\hat{\sigma}\right]
\end{equation}
in terms of the change they induce on \(\hat{\sigma}\) and \(\hat{j}\). In terms of these, the remaining scalar components obey the transformation
\begin{gather}
\delta_X\hat{\xi} = -\mathscr{A}\delta_X\hat{\sigma}\,, \quad
\delta_X\hat{y} = \frac{A\partial_t\delta_X\hat{\sigma}}{N} + \left(\frac{\partial_tA}{N} - \mathscr{V}\right)\delta_X\hat{\sigma}\,, \quad
\delta_X\hat{\psi} = -\frac{\partial_tA}{N}\left[\delta_X\hat{j} + \left(\frac{\partial_tA}{N} - \mathscr{V}\right)\delta_X\hat{\sigma}\right]\,,\nonumber\\
\delta_X\hat{\phi} = \frac{\partial_tA}{N}\left[\delta_X\hat{j} + \left(\frac{\partial_tA}{N} - \mathscr{V}\right)\delta_X\hat{\sigma}\right] + \frac{A}{N}\partial_t\left[\delta_X\hat{j} + \left(\frac{\partial_tA}{N} - \mathscr{V}\right)\delta_X\hat{\sigma}\right]\,.
\end{gather}
From these immediately follow the gauge-invariant combinations
\begin{gather}
\hat{\boldsymbol{\xi}} = \hat{\xi} + \mathscr{A}\hat{\sigma}\,, \quad
\hat{\mathbf{y}} = \hat{y} - \frac{A\partial_t\hat{\sigma}}{N} - \left(\frac{\partial_tA}{N} - \mathscr{V}\right)\hat{\sigma}\,, \quad
\hat{\boldsymbol{\psi}} = \hat{\psi} + \frac{\partial_tA}{N}\left[\hat{j} + \left(\frac{\partial_tA}{N} - \mathscr{V}\right)\hat{\sigma}\right]\,,\nonumber\\
\hat{\boldsymbol{\phi}} = \hat{\phi} - \frac{\partial_tA}{N}\left[\hat{j} + \left(\frac{\partial_tA}{N} - \mathscr{V}\right)\hat{\sigma}\right] + \frac{A}{N}\partial_t\left[\hat{j} + \left(\frac{\partial_tA}{N} - \mathscr{V}\right)\hat{\sigma}\right]\,.
\end{gather}
As for the vector components, also these are most conveniently expressed using conformal time and the conformal Hubble parameter, which yields the expressions
\begin{gather}
\hat{\boldsymbol{\xi}} = \hat{\xi} + \mathscr{A}\hat{\sigma}\,, \quad
\hat{\mathbf{y}} = \hat{y} - \hat{\sigma}' - (\mathcal{H} - \mathscr{V})\hat{\sigma}\,, \quad
\hat{\boldsymbol{\psi}} = \hat{\psi} + \mathcal{H}[\hat{j} + (\mathcal{H} - \mathscr{V})\hat{\sigma}]\,,\nonumber\\
\hat{\boldsymbol{\phi}} = \hat{\phi} - \mathcal{H}[\hat{j} + (\mathcal{H} - \mathscr{V})\hat{\sigma}] + [\hat{j} + (\mathcal{H} - \mathscr{V})\hat{\sigma}]'\,.
\end{gather}
These are also easily obtained from the gauge transformations of the perturbations~\eqref{eq:gtrandecct}.

We see that both for the vector and the scalar components the gauge-invariant combinations of the general cosmological tetrad perturbation receive additional contributions which are not present in the spatially flat background~\eqref{eq:tetfuncflat}. In particular, we see that the parity-odd gauge-invariant perturbations \(\hat{\boldsymbol{\xi}}\) and \(\hat{\mathbf{w}}_a\) depend on the parameter \(\mathscr{A}\), which is non-vanishing only for the axial tetrad~\eqref{eq:tetfuncaxi}. The parity-even components, in contrast, receive contributions from the term \(\mathscr{V} - \mathcal{H}\), which is non-vanishing only for the vector branch~\eqref{eq:tetfuncvec}.

\section{Application to teleparallel gravity}\label{sec:gravity}
We finally discuss how the theory of cosmological perturbations of teleparallel geometry can be applied to gravity theories which are built upon this geometry. For this purpose, we briefly review the general structure of the action and field equations of teleparallel gravity theories in section~\ref{ssec:feqstruc}, and discuss the general form of their perturbations around a cosmologically symmetric background. We then show how these can be expressed in terms of the perturbations of the teleparallel geometry, which we derived in section~\ref{ssec:pertele}, in section~\ref{ssec:feqpert}. This is further simplified by making once again use of the irreducible decomposition approach in section~\ref{ssec:feqdec}. We then discuss their behavior under gauge transformations in section~\ref{ssec:feqgauge}. Finally, in section~\ref{ssec:tegr} we demonstrate these findings by applying them to the well-known example of TEGR.

\subsection{Structure of the perturbed field equations}\label{ssec:feqstruc}
In the following we will assume that the action of the teleparallel gravity theory under consideration can be decomposed in the form
\begin{equation}
S[\theta, \omega, \Omega] = S_{\text{g}}[\theta, \omega] + S_{\text{m}}[\theta, \Omega]\,,
\end{equation}
where the gravitational part \(S_{\text{g}}\) depends only on the tetrad and the spin connection, while the matter part \(S_{\text{m}}\) depends only on the tetrad and an arbitrary set \(\Omega\) of matter fields. Writing the variation of the action with respect to the tetrad in the form
\begin{equation}
\delta_{\theta}S_{\text{g}} = \int_M\delta\theta^A{}_{\mu}E_A{}^{\mu}\theta\dd^4x\,, \quad
\delta_{\theta}S_{\text{m}} = -\int_M\delta\theta^A{}_{\mu}\Theta_A{}^{\mu}\theta\dd^4x\,,
\end{equation}
where \(\theta\) denotes the determinant of the tetrad, the field equations obtained from variation with respect to the tetrad can be written as
\begin{equation}\label{eq:telefield}
E_A{}^{\mu} = \Theta_A{}^{\mu}\,.
\end{equation}
A few more statements can be made by transforming the appearing indices on the Euler-Lagrange expression \(E_A{}^{\mu}\) and the energy-momentum tensor \(\Theta_A{}^{\mu}\) into lower spacetime indices, using the tetrad and the metric. We further assume that the matter fields are minimally coupled to the metric \(g_{\mu\nu}\) only. It then follows that the matter action \(S_{\text{m}}\) is invariant under local Lorentz transformations, which further implies that the energy-momentum tensor is symmetric, \(\Theta_{[\mu\nu]} = 0\). We make a similar assumption also for the gravitational part \(S_{\text{g}}\). Here local Lorentz invariance follows from the assumption that the action is constructed from the metric \(g_{\mu\nu}\) and the torsion tensor \(\tp{T}^{\mu}{}_{\nu\rho}\) only. It then follows that the Euler-Lagrange expressions derived by variation of the spin connection, while maintaining its flatness and metric compatibility, are exactly the antisymmetric part of the tetrad field equations~\cite{Hohmann:2017duq}. The field equations thus split into a symmetric and antisymmetric part, which take the general form
\begin{equation}
E_{(\mu\nu)} = \Theta_{\mu\nu}\,, \quad
E_{[\mu\nu]} = 0\,.
\end{equation}
Further assuming that the gravitational part of the action does not contain any derivatives acting on the torsion or the metric, so that there are no derivatives acting on the dynamical fields aside from the first-order derivative of the tetrad which enters the torsion, implies that the tensor \(E_{\mu\nu}\) is fully expressed in terms of the metric, the torsion and first-order covariant derivatives acting on the torsion.

We then consider the case that the tetrad is given in the form~\eqref{eq:tetpert} as a perturbation around a cosmologically symmetric tetrad, while the spin connection vanishes following our choice to work in the Weitzenböck gauge. Applying this to the field equations written in the form~\eqref{eq:telefield}, it follows that also their gravitational and matter parts split in the form
\begin{equation}\label{eq:fieldpert}
E_A{}^{\mu} = \bar{E}_A{}^{\mu} + \mathcal{E}_A{}^{\mu}\,, \quad
\Theta_A{}^{\mu} = \bar{\Theta}_A{}^{\mu} + \mathcal{T}_A{}^{\mu}
\end{equation}
into a cosmologically symmetric background \(\bar{E}_A{}^{\mu}\), \(\bar{\Theta}_A{}^{\mu}\) and a perturbation \(\mathcal{E}_A{}^{\mu}\), \(\mathcal{T}_A{}^{\mu}\). The background satisfies the cosmological field equations
\begin{equation}
\bar{E}_A{}^{\mu} = \bar{\Theta}_A{}^{\mu}\,,
\end{equation}
where cosmological symmetry mandates that the background of the energy-momentum tensor takes the perfect fluid form
\begin{equation}\label{eq:enmomcosmo}
\bar{\Theta}_{\mu\nu} = (\bar{\rho} + \bar{p})n_{\mu}n_{\nu} + \bar{p}g_{\mu\nu} = \bar{\rho}n_{\mu}n_{\nu} + \bar{p}h_{\mu\nu}
\end{equation}
with matter density \(\bar{\rho}\) and pressure \(\bar{p}\), while the geometry part takes the same form
\begin{equation}\label{eq:gravcosmo}
\bar{E}_{\mu\nu} = \mathfrak{N}n_{\mu}n_{\nu} + \mathfrak{H}h_{\mu\nu}\,,
\end{equation}
with components \(\mathfrak{N}\) and \(\mathfrak{H}\) which are determined in terms of the teleparallel geometry by the particular gravity theory under consideration. The background field equations thus reduce to the simple form
\begin{equation}\label{eq:background}
\mathfrak{N} = \bar{\rho}\,, \quad
\mathfrak{H} = \bar{p}\,.
\end{equation}
See~\cite{Hohmann:2020zre} for an exhaustive discussion and several examples. In the following, we will assume that these equations are satisfied, and so we can focus on the perturbation equations
\begin{equation}
\mathcal{E}_A{}^{\mu} = \mathcal{T}_A{}^{\mu}\,,
\end{equation}
which will determine the solution for the tetrad perturbation.

\subsection{Perturbation of the gravity part}\label{ssec:feqpert}
We now turn our attention to the perturbation of the field equations. For the perturbation \(\mathcal{T}_{\mu\nu} = \bar{\theta}^A{}_{\mu}g_{\nu\rho}\mathcal{T}_A{}^{\rho}\) of the energy-momentum tensor we assume the standard form used in cosmological perturbation theory; see, e.g.,~\cite{Mukhanov:1990me} for a comprehensive review. Here we will restrict our focus to the gravitational part \(\mathcal{E}_{\mu\nu}\) of the perturbed field equations. Following our assumption that \(E_{\mu\nu}\) is composed from the metric, the torsion and first-order covariant derivatives of the torsion, it follows that a linear perturbation can uniquely be written in the form
\begin{equation}
\mathcal{E}_{\mu\nu} = F^{(1)}_{\mu\nu}{}^{\alpha\beta}\delta g_{\alpha\beta} + F^{(2)}_{\mu\nu\alpha}{}^{\beta\gamma}\delta\tp{T}^{\alpha}{}_{\beta\gamma} + F^{(3)}_{\mu\nu\alpha}{}^{\beta\gamma\delta}\delta\tp{\nabla}_{\delta}\tp{T}^{\alpha}{}_{\beta\gamma}\,,
\end{equation}
where the three tensors \(F^{(i)}\) are composed of the metric, torsion and first-order covariant derivative of the torsion, evaluated at the cosmological background. It follows that they obey the symmetry condition,
\begin{equation}
\mathcal{L}_XF^{(i)} = 0\,,
\end{equation}
where \(X^{\mu}\) is any of the generators~\eqref{eq:genrot} and~\eqref{eq:gentra} of the cosmological symmetry. It follows from this fact that the most general perturbation \(\mathcal{E}_{\mu\nu}\) has a rather simple structure. First, it can be obtained from the tetrad perturbation~\eqref{eq:perttens} by calculating the perturbations \(\delta g_{\alpha\beta}, \delta\tp{T}^{\alpha}{}_{\beta\gamma}, \delta\tp{\nabla}_{\delta}\tp{T}^{\alpha}{}_{\beta\gamma}\) displayed in section~\ref{ssec:pertele}. Second, from the cosmological symmetry of the contraction tensors \(F^{(i)}\) one can derive their most general form, and it turns out that each of them can be expressed as a finite linear combination of tensors composed from the background objects \(n_{\mu}, h_{\mu\nu}, \varepsilon_{\mu\nu\rho}\), with coefficients which depend only on the time coordinate \(t\) and which depend on the gravity theory under consideration. Taking into account the symmetries
\begin{equation}
F^{(1)}_{\mu\nu}{}^{[\alpha\beta]} = 0\,, \quad
F^{(2)}_{\mu\nu\alpha}{}^{(\beta\gamma)} = 0\,, \quad
F^{(3)}_{\mu\nu\alpha}{}^{(\beta\gamma)\delta} = 0\,,
\end{equation}
one finds that imposing cosmological symmetry restricts \(F^{(1)}\) to a linear combination of 8 terms, \(F^{(2)}\) has 18 terms and \(F^{(3)}\) contains 28 terms. While it is in principle possible to list all these terms and calculate the corresponding contributions to the perturbed field equations, it turns out that one may follow a simpler approach, which is not less general.

\subsection{Irreducible decomposition}\label{ssec:feqdec}
To further simplify the perturbed field equations, we perform an irreducible decomposition, in analogy to the decomposition of the tetrad perturbations shown in section~\ref{ssec:perdec}. For the perturbation of the field equations we define them as
\begin{equation}
\mathcal{E}_{\mu\nu} = \Phi n_{\mu}n_{\nu} - (B_{\nu} + A\lc{\DD}_{\nu}J)n_{\mu} - (V_{\mu} + A\lc{\DD}_{\mu}Y)n_{\nu} + \Psi h_{\mu\nu} + A^2\lc{\DD}_{\mu}\lc{\DD}_{\nu}\Sigma + A\lc{\DD}_{\mu}C_{\nu} + \varepsilon_{\mu\nu\rho}(A\lc{\DD}^{\rho}\Xi + W^{\rho}) + \frac{1}{2}Q_{\mu\nu}\,,
\end{equation}
where, as for the tetrad perturbations, the vectors \(B_{\mu}, V_{\mu}, C_{\mu}, W_{\mu}\) and the symmetric, trace-free tensor \(Q_{\mu\nu}\) are spatial and divergence-free. After performing a split into time and space components, these take the form
\begin{equation}\label{eq:feqdec}
\hat{\mathcal{E}}_{00} = \hat{\Phi}\,, \quad
\hat{\mathcal{E}}_{0b} = \dd_b\hat{J} + \hat{B}_b\,, \quad
\hat{\mathcal{E}}_{a0} = \dd_a\hat{Y} + \hat{V}_a\,, \quad
\hat{\mathcal{E}}_{ab} = \hat{\Psi}\gamma_{ab} + \dd_a\dd_b\Sigma + \dd_a\hat{C}_b + \upsilon_{abc}(\dd^c\hat{\Xi} + \hat{W}^c) + \frac{1}{2}\hat{Q}_{ab}\,.
\end{equation}
Note that similarly to the tetrad perturbation, we do not symmetrize the term \(\dd_a\hat{C}_b\) in the purely spatial part of the perturbations. However, he we make the crucial distinction that the order of the indices is reversed compared to the similar term \(\dd_b\hat{c}_a\) in the tetrad perturbation. The reason for this opposite order of the indices will become clear when we discuss how these components transform under infinitesimal coordinate changes below.

The irreducible decomposition now significantly simplifies the task of determining the cosmological perturbation for the general class of teleparallel gravity theories, compared to our considerations mentioned in section~\ref{ssec:feqpert}. Instead of constructing the perturbation \(\mathcal{E}_{\mu\nu}\) from the tetrad perturbation \(\tau_{\mu\nu}\) and its (spacetime) covariant derivatives, the task is broken down to calculating the irreducible components which appear in the decomposition~\eqref{eq:feqdec}. It is the virtue of the irreducible decomposition that the scalar, vector and tensor components decouple, and can therefore be studied separately.

\subsection{Gauge transformations and gauge invariance}\label{ssec:feqgauge}
We finally study the behavior of the perturbed teleparallel field equations under infinitesimal coordinate transformations of the form~\eqref{eq:coordtrans}. We proceed in analogy to section~\ref{sec:gauge}, where we have studied gauge transformations of the tetrad perturbations. The starting point will be the field equations~\eqref{eq:telefield}, together with the perturbation ansatz~\eqref{eq:fieldpert}. Considering that an infinitesimal change of the field equations is given by the Lie derivative
\begin{equation}
E_A{}^{\mu} = E'_A{}^{\mu} + (\mathcal{L}_X\bar{E})_A{}^{\mu}\,,
\end{equation}
of the unperturbed field equations, if we restrict ourselves to terms linear in any perturbation. We then express the transformed field equations as a perturbation of the same unperturbed background,
\begin{equation}
E'_A{}^{\mu} = \bar{E}_A{}^{\mu} + \mathcal{E}'_A{}^{\mu}\,.
\end{equation}
It then follows that the transformation of the perturbation is, analogously to the tetrad perturbation, given by
\begin{equation}\label{eq:feqlieder}
\delta_X\mathcal{E}_A{}^{\mu} = \mathcal{E}_A{}^{\mu} - \mathcal{E}'_A{}^{\mu} = (\mathcal{L}_X\bar{E})_A{}^{\mu} = X^{\nu}\partial_{\nu}\bar{E}_A{}^{\mu} - \partial_{\nu}X^{\mu}\bar{E}_A{}^{\nu}\,.
\end{equation}
We then transform the indices with the background geometry, in order to write the gauge transformation in the form
\begin{equation}
\delta_X\mathcal{E}_{\mu\nu} = \bar{\theta}^A{}_{\mu}\bar{g}_{\nu\rho}\delta_X\mathcal{E}_A{}^{\rho}\,.
\end{equation}
Inserting the form~\eqref{eq:gravcosmo} of the background equations, this can now further be decomposed into space and time components. Using the decomposition~\eqref{eq:gaugedec} of the gauge transformation vector field, the transformation of the field equations splits into the components
\begin{subequations}
\begin{align}
\delta_X\hat{\mathcal{E}}_{00} &= \frac{\mathfrak{N}\partial_t\hat{X}_{\perp} - \hat{X}_{\perp}\partial_t\mathfrak{N}}{N}\,,\\
\delta_X\hat{\mathcal{E}}_{0b} &= \frac{[N\mathscr{V}\mathfrak{H} - (\mathfrak{N} + \mathfrak{H})\partial_tA](\dd_b\hat{X}_{\parallel} + \hat{Z}_b) + \mathfrak{N}A\partial_t(\dd_b\hat{X}_{\parallel} + \hat{Z}_b)}{NA}\,,\\
\delta_X\hat{\mathcal{E}}_{a0} &= \frac{(N\mathscr{V} - \partial_tA)\mathfrak{N}(\dd_a\hat{X}_{\parallel} + \hat{Z}_a) - N\mathfrak{H}\dd_a\hat{X}_{\perp}}{NA}\,,\\
\delta_X\hat{\mathcal{E}}_{ab} &= \frac{(\mathfrak{H}\partial_tA - A\partial_t\mathfrak{H})\hat{X}_{\perp}\gamma_{ab} - N\mathfrak{H}\mathscr{A}\upsilon_{abc}(\dd^c\hat{X}_{\parallel} + \hat{Z}^c) - N\mathfrak{H}\dd_a(\dd_b\hat{X}_{\parallel} + \hat{Z}_b)}{NA}\,.
\end{align}
\end{subequations}
In terms of conformal time and the conformal Hubble parameter, we find the simpler formulas
\begin{subequations}
\begin{align}
A\delta_X\hat{\mathcal{E}}_{00} &= \mathfrak{N}\hat{X}_{\perp}' - \hat{X}_{\perp}\mathfrak{N}'\,,\\
A\delta_X\hat{\mathcal{E}}_{0b} &= [(\mathscr{V} - \mathcal{H})\mathfrak{H} - \mathcal{H}\mathfrak{N}](\dd_b\hat{X}_{\parallel} + \hat{Z}_b) + \mathfrak{N}(\dd_b\hat{X}_{\parallel}' + \hat{Z}_b')\,,\\
A\delta_X\hat{\mathcal{E}}_{a0} &= (\mathscr{V} - \mathcal{H})\mathfrak{N}(\dd_a\hat{X}_{\parallel} + \hat{Z}_a) - \mathfrak{H}\dd_a\hat{X}_{\perp}\,,\\
A\delta_X\hat{\mathcal{E}}_{ab} &= (\mathfrak{H}\mathcal{H} - \mathfrak{H}')\hat{X}_{\perp}\gamma_{ab} - \mathfrak{H}\mathscr{A}\upsilon_{abc}(\dd^c\hat{X}_{\parallel} + \hat{Z}^c) - \mathfrak{H}\dd_a(\dd_b\hat{X}_{\parallel} + \hat{Z}_b)\,.
\end{align}
\end{subequations}
Comparing this result with the decomposition~\eqref{eq:feqdec} of the field equations, we find that their irreducible components transform as
\begin{gather}
\delta_X\hat{V}_a = \frac{(N\mathscr{V} - \partial_tA)\mathfrak{N}\hat{Z}_a}{NA}\,, \quad
\delta_X\hat{\Phi} = \frac{\mathfrak{N}\partial_t\hat{X}_{\perp} - \hat{X}_{\perp}\partial_t\mathfrak{N}}{N}\,, \quad
\delta_X\hat{\Psi} = \frac{\mathfrak{H}\partial_tA - A\partial_t\mathfrak{H}}{NA}\hat{X}_{\perp}\,, \quad
\delta_X\hat{\Sigma} = -\frac{\mathfrak{H}\hat{X}_{\parallel}}{A}\,,\nonumber\\
\delta_X\hat{\Xi} = -\frac{\mathfrak{H}\mathscr{A}\hat{X}_{\parallel}}{A}\,, \quad
\delta_X\hat{J} = \frac{[N\mathscr{V}\mathfrak{H} - (\mathfrak{N} + \mathfrak{H})\partial_tA]\hat{X}_{\parallel} + \mathfrak{N}A\partial_t\hat{X}_{\parallel}}{NA}\,, \quad
\delta_X\hat{Y} = \frac{(N\mathscr{V} - \partial_tA)\mathfrak{N}\hat{X}_{\parallel} - N\mathfrak{H}\hat{X}_{\perp}}{NA}\,,\nonumber\\
\delta_X\hat{C}_a = -\frac{\mathfrak{H}\hat{Z}_a}{A}\,, \quad
\delta_X\hat{W}_a = -\frac{\mathfrak{H}\mathscr{A}\hat{Z}_a}{A}\,, \quad
\delta_X\hat{B}_a = \frac{[N\mathscr{V}\mathfrak{H} - (\mathfrak{N} + \mathfrak{H})\partial_tA]\hat{Z}_a + \mathfrak{N}A\partial_t\hat{Z}_a}{NA}\,, \quad
\delta_X\hat{Q}_{ab} = 0\,,
\end{gather}
which translates to
\begin{gather}
A\delta_X\hat{V}_a = (\mathscr{V} - \mathcal{H})\mathfrak{N}\hat{Z}_a\,, \quad
A\delta_X\hat{\Phi} = \mathfrak{N}\hat{X}_{\perp}' - \hat{X}_{\perp}\mathfrak{N}'\,, \quad
A\delta_X\hat{\Psi} = (\mathfrak{H}\mathcal{H} - \mathfrak{H}')\hat{X}_{\perp}\,, \quad
A\delta_X\hat{\Sigma} = -\mathfrak{H}\hat{X}_{\parallel}\,,\nonumber\\
A\delta_X\hat{\Xi} = -\mathfrak{H}\mathscr{A}\hat{X}_{\parallel}\,, \quad
A\delta_X\hat{J} = [(\mathscr{V} - \mathcal{H})\mathfrak{H} - \mathcal{H}\mathfrak{N}]\hat{X}_{\parallel} + \mathfrak{N}\hat{X}_{\parallel}'\,, \quad
A\delta_X\hat{Y} = (\mathscr{V} - \mathcal{H})\mathfrak{N}\hat{X}_{\parallel} - \mathfrak{H}\hat{X}_{\perp}\,,\nonumber\\
A\delta_X\hat{C}_a = -\mathfrak{H}\hat{Z}_a\,, \quad
A\delta_X\hat{W}_a = -\mathfrak{H}\mathscr{A}\hat{Z}_a\,, \quad
A\delta_X\hat{B}_a = [(\mathscr{V} - \mathcal{H})\mathfrak{H} - \mathcal{H}\mathfrak{N}]\hat{Z}_a + \mathfrak{N}\hat{Z}_a'\,, \quad
A\delta_X\hat{Q}_{ab} = 0
\end{gather}
in the conformal parametrization. Here we can easily identify the transformation of the component \(\hat{C}_a\) due to the choice of the order of indices in the expansion~\eqref{eq:feqdec}. This change is the order of the indices, compared to the tetrad perturbation~\eqref{eq:pertdec}, arises from the order of the indices in the term \(\partial_{\alpha}X^{\beta}\) in the Lie derivatives~\eqref{eq:pertlieder} and~\eqref{eq:feqlieder}.

Note that, by construction, the result is linear in the terms \(\mathfrak{N}\) and \(\mathfrak{H}\) which constitute the geometry part of the field equations for the cosmologically symmetric background. It follows further that performing a gauge transformation of the energy-momentum tensor \(\Theta_A{}^{\mu}\) yields the same form, where the matter density \(\bar{\rho}\) and pressure \(\bar{p}\) appear in place of \(\mathfrak{N}\) and \(\mathfrak{H}\). Once the background field equations~\eqref{eq:background} are imposed, the gauge transformations of both sides of the field equations cancel each other, so that only gauge-invariant quantities remain.

\subsection{Example: TEGR}\label{ssec:tegr}
As an illustrative example, we study the teleparallel equivalent of gravity (TEGR), whose gravitational part of the action is given by~\cite{Maluf:2013gaa}
\begin{equation}\label{eq:tegraction}
S_{\text{g}} = \frac{1}{2\kappa^2}\int\dd^4x\,\theta\,\mathbb{T}
\end{equation}
in terms of the torsion scalar
\begin{equation}\label{eq:torsscal}
\mathbb{T} = \frac{1}{2}\tp{T}^{\rho}{}_{\mu\nu}\tp{S}_{\rho}{}^{\mu\nu}\,,
\end{equation}
where we made use of the superpotential
\begin{equation}\label{eq:suppot}
\tp{S}_{\rho}{}^{\mu\nu} = \tp{K}^{\mu\nu}{}_{\rho} - \delta_{\rho}^{\mu}\tp{T}_{\sigma}{}^{\sigma\nu} + \delta_{\rho}^{\nu}\tp{T}_{\sigma}{}^{\sigma\mu}\,.
\end{equation}
Variation with respect to the tetrad yields the gravitational part of the field equations
\begin{equation}\label{eq:tegrfeq}
\kappa^2E_{\mu\nu} = \frac{1}{2}\mathbb{T}g_{\mu\nu} + \lc{\nabla}_{\rho}\left(\tp{S}_{\nu\mu}{}^{\rho}\right) + \tp{S}^{\rho\sigma}{}_{\mu}\left(\tp{K}_{\rho\nu\sigma} - \tp{T}_{\rho\sigma\nu}\right)\,.
\end{equation}
For the discussion of the cosmology, we now switch to the conformal parametrization for simplicity. First, note that the geometry side of the background equations is given by
\begin{equation}\label{eq:tegrback}
\kappa^2\mathfrak{N} = 3\left(H^2 + \frac{u^2}{A^2}\right) = \frac{3}{A^2}(\mathcal{H}^2 + u^2)\,, \quad
\kappa^2\mathfrak{H} = -2\dot{H} - 3H^2 - \frac{u^2}{A^2} = -\frac{1}{A^2}(2\mathcal{H}' + \mathcal{H}^2 + u^2)\,.
\end{equation}
We then follow the procedure outlined in the previous sections to derive the perturbed field equations. In the decomposition~\eqref{eq:feqdec}, we find that the perturbation of the geometry side takes the form
\begin{subequations}
\begin{align}
\kappa^2\hat{\Phi} &= 3(3\mathcal{H}^2 + u^2)\hat{\phi} + 2\mathcal{H}\triangle(\hat{\sigma}' - \hat{j} - \hat{y}) + 6\mathcal{H}\hat{\psi}' - 6u^2\hat{\psi} - 2\triangle\hat{\psi}\,,\\
\kappa^2\hat{\Psi} &= \triangle\left[\hat{\psi} - \hat{\phi} + 2\mathcal{H}(\hat{j} + \hat{y} - \hat{\sigma}') + \hat{j}' + \hat{y}' - \hat{\sigma}''\right] + (\mathcal{H}^2 + 2\mathcal{H}' + 3u^2)\hat{\psi} - 4\mathcal{H}\hat{\psi}' - 2\hat{\psi}'' - 2(\mathcal{H}^2 + 2\mathcal{H}')\hat{\phi} - 2\mathcal{H}\hat{\phi}'\,,\\
\kappa^2\hat{\Xi} &= -(\mathcal{H}^2 + 2\mathcal{H}' + u^2)\hat{\xi}\,,\\
\kappa^2\hat{\Sigma} &= \hat{\phi} - \hat{\psi} - 2\mathcal{H}(\hat{j} + \hat{y} - \hat{\sigma}') - \hat{j}' - \hat{y}' + (\mathcal{H}^2 + 2\mathcal{H}' + u^2)\hat{\sigma} + \hat{\sigma}''\,,\\
\kappa^2\hat{J} &= 2u^2\hat{\sigma}' - 2\hat{\psi}' + (3\mathcal{H}^2 + u^2)\hat{y} + 2(\mathcal{H}^2 - \mathcal{H}')\hat{j} - 2\mathcal{H}\hat{\phi}\,,\\
\kappa^2\hat{Y} &= 2u^2\hat{\sigma}' - 2\hat{\psi}' - 2u^2\hat{y} + (3\mathcal{H}^2 + u^2)\hat{j} - 2\mathcal{H}\hat{\phi}\,,\\
\kappa^2\hat{C}_a &= \hat{c}_a'' - \hat{b}_a' - \hat{v}_a' + 2\mathcal{H}(\hat{c}_a' - \hat{b}_a - \hat{v}_a) + (\mathcal{H}^2 + 2\mathcal{H}' + u^2)\hat{c}_a\,,\\
\kappa^2\hat{W}_a &= \upsilon_{abc}\dd^b\left[\frac{1}{2}({\hat{c}^c}'' - {\hat{b}^c}' - {\hat{v}^c}') + \mathcal{H}({\hat{c}^c}' - \hat{b}^c - \hat{v}^c)\right] - (\mathcal{H}^2 + 2\mathcal{H}' + u^2)\hat{w}_a\,,\\
\kappa^2\hat{B}_a &= 2(\mathcal{H}^2 - \mathcal{H}')\hat{b}_a + 3\mathcal{H}^2\hat{v} + u^2(\hat{b}_a + 2\hat{v}_a + \hat{c}_a') - \frac{1}{2}\triangle(\hat{b}_a + \hat{v}_a - \hat{c}_a')\,,\\
\kappa^2\hat{V}_a &= 3\mathcal{H}^2\hat{b}_a + u^2(2\hat{b}_a - \hat{v}_a + \hat{c}_a') - \frac{1}{2}\triangle(\hat{b}_a + \hat{v}_a - \hat{c}_a')\,,\\
\kappa^2\hat{Q}_{ab} &= \hat{q}_{ab}'' + 2\mathcal{H}\hat{q}_{ab}' - \triangle\hat{q}_{ab} + (3u^2 + \mathcal{H}^2 + 2\mathcal{H}')\hat{q}_{ab}\,.
\end{align}
\end{subequations}
It turns out that these are independent of the functions \(\mathscr{V}\) and \(\mathscr{A}\) which distinguish between the different cosmologically symmetric background tetrads, since only the metric degrees of freedom enter the field equations. Further, one sees that the resulting equations \(\mathcal{E}_{\mu\nu}\) are not symmetric in their two indices, although the TEGR field equations~\eqref{eq:tegrfeq}, which are equivalent to Einstein's equations of general relativity, are symmetric. This is due to our definition~\eqref{eq:fieldpert} of the perturbed field equations \(E_A{}^{\mu}\) with mixed indices. It follows that in addition to the perturbation of the symmetric field equations \(E_{\mu\nu}\) one receives a perturbation of the tetrad \(\theta^A{}_{\mu}\) which is used to transform the indices. This perturbation, which possesses an antisymmetric part \(\tau_{[\mu\nu]}\), is proportional to the background field equations \(\bar{E}_{\mu\nu}\). Indeed, we find that the antisymmetric part, which is governed by the expressions
\begin{subequations}
\begin{align}
\kappa^2\hat{\Xi} &= -(\mathcal{H}^2 + 2\mathcal{H}' + u^2)\hat{\xi}\,,\\
\kappa^2(\hat{J} - \hat{Y}) &= 3(\mathcal{H}^2 + u^2)\hat{y} - (\mathcal{H}^2 + 2\mathcal{H}' + u^2)\hat{j}\,,\\
\kappa^2(\hat{B}_a - \hat{V}_a) &= 3(\mathcal{H}^2 + u^2)\hat{v}_a - (\mathcal{H}^2 + 2\mathcal{H}' + u^2)\hat{b}_a\,,\\
\kappa^2(\dd_{[a}\hat{C}_{b]} + \upsilon_{abc}\hat{W}^c) &= (\mathcal{H}^2 + 2\mathcal{H}' + u^2)(\dd_{[a}\hat{c}_{b]} - \upsilon_{abc}\hat{w}^c)\,,
\end{align}
\end{subequations}
is proportional to a linear combination of the background equations~\eqref{eq:tegrback}, and cancels with a corresponding contribution from the energy-momentum tensor once the background field equations are imposed. For the remaining symmetric part of the perturbation equations, one finds after imposing the background equations, that these resemble the perturbative expansion of Einstein's field equations, up to a redefinition of the metric perturbations in terms of the tetrad~\cite{Mukhanov:1990me}.

\section{Conclusion}\label{sec:conclusion}
We have studied perturbations of the most general cosmologically symmetric (homogeneous and isotropic) teleparallel geometries. Working in the Weitzenböck gauge, where the spin connection is imposed to vanish at all perturbation orders, these are given by perturbations of the tetrad only. We have decomposed these tetrad perturbations into irreducible components under the action of the spatial rotation group. Further, we have studied how these components behave under gauge transformations, i.e., infinitesimal coordinate transformations which retain the order of magnitude of the tetrad perturbations. From these transformations we have obtained gauge-invariant quantities.

Moreover, we have discussed the application of these perturbations to teleparallel gravity theories. For this purpose we have studied the general structure of the field equations of a teleparallel gravity theory for a perturbation around a cosmologically symmetric background, their decomposition into irreducible components and their transformation under infinitesimal coordinate changes. To illustrate the formalism, we have applied it to derive the perturbative expansion of the field equations for the teleparallel equivalent of general relativity (TEGR).

Our work lays the foundation to study the evolution of perturbations of teleparallel gravity theories around general cosmologically symmetric teleparallel background spacetimes, and thus to extend previous studies which assume a spatially flat diagonal background tetrad. We have seen that the perturbative expansion of the field equations in general depends on the choice of the cosmological background tetrad~\cite{Hohmann:2020zre}, which enters via two scalar functions measuring its vector and axial torsion components. One may therefore expect that they will lead to a different evolution of the perturbations, possibly with observable effects, which may be used to constrain modified teleparallel theories. It may also lead to new insights to the question about the presence or absence of additional perturbative degrees of freedom around different backgrounds~\cite{Ferraro:2011us,Ferraro:2014owa,Ferraro:2018axk,Bejarano:2019fii,Ferraro:2020tqk,Jimenez:2020ofm}.

Another possibility for future work is to extend the formalism we presented here to higher order perturbations. For cosmologically symmetric Riemannian spacetimes, there exists a well-established theory~\cite{Tomita:PTP.37.831,Tomita:PTP.45.1747,Tomita:PTP.47.416}, as well as a gauge-invariant approach~\cite{Nakamura:2004rm,Nakamura:2006rk}. The underlying gauge-invariant higher order perturbation theory~\cite{Bruni:1996im,Bruni:1999et,Sonego:1997np} has already been successfully applied to teleparallel geometry in the context of the parametrized post-Newtonian formalism~\cite{Hohmann:2019qgo}, where a perturbation around Minkowski spacetime is assumed, and so can serve as the foundation to derive also higher-order cosmological perturbations of teleparallel spacetime geometry.

\begin{acknowledgement}
The author gratefully acknowledges the full support by the Estonian Research Council through the Personal Research Funding project PRG356, as well as the European Regional Development Fund through the Center of Excellence TK133 ``The Dark Side of the Universe''.
\end{acknowledgement}

\bibliography{pertrefs}
\bibliographystyle{plainurl}
\end{document}